\documentclass[aps, prb, reprint, superscriptaddress]{revtex4-2}
\usepackage{amsmath, amssymb, graphicx, hyperref, csquotes}

\newcommand{\gk}{$\bar{\mathit{\varGamma}}\!\bar{K}$}
\newcommand{\gm}{$\bar{\mathit{\varGamma}}\!\bar{M}$}
\newcommand{\kgk}{$\bar{K}\!\bar{\mathit{\varGamma}}\!\bar{K}'$}

\newcommand{\mbt}{$\mathrm{MnBi}_2 \mathrm{Te}_4$}

\newcommand{\gbt}{$\mathrm{GeBi}_2 \mathrm{Te}_4$}
\newcommand{\mabt}{$\mathrm{Mn}_{1-x} A_x \mathrm{Bi}_2 \mathrm{Te}_4$}
\newcommand{\mgbt}{$\mathrm{Mn}_{1-x} \mathrm{Ge}_x \mathrm{Bi}_2 \mathrm{Te}_4$}
\newcommand{\mgbttitle}{\texorpdfstring{$\mathbf{Mn}_{1-x} \mathbf{Ge}_x \mathbf{Bi}_2 \mathbf{Te}_4$}{Mn(1-x)Ge(x)Bi2Te4}}
\newcommand{\msbt}{$\mathrm{Mn}_{1-x} \mathrm{Sn}_x \mathrm{Bi}_2 \mathrm{Te}_4$}

\begin{document}

\title{Spin texture tunability in \mgbttitle{} through varying Ge Concentration
}

\author{A.M.~Shikin}
    \email{ashikin@inbox.ru}
    \affiliation{Saint Petersburg State University, 198504 Saint Petersburg, Russia}
\author{D.A.~Estyunin}
    \affiliation{Saint Petersburg State University, 198504 Saint Petersburg, Russia}
\author{N.L.~Zaitsev}
    \affiliation{Saint Petersburg State University, 198504 Saint Petersburg, Russia}
    \affiliation{Institute of Molecule and Crystal Physics, Subdivision of the Ufa Federal Research Centre of the Russian Academy of Sciences, 450075 Ufa, Russia}
\author{T.P.~Estyunina}
    \affiliation{Saint Petersburg State University, 198504 Saint Petersburg, Russia}
\author{A.V.~Eryzhenkov}
    \affiliation{Saint Petersburg State University, 198504 Saint Petersburg, Russia}
\author{K.A.~Kokh}
    \affiliation{Saint Petersburg State University, 198504 Saint Petersburg, Russia}
    \affiliation{Sobolev Institute of Geology and Mineralogy, Siberian Branch, Russian Academy of Sciences, 630090 Novosibirsk, Russia}
\author{O.E.~Tereshchenko}
    \affiliation{Saint Petersburg State University, 198504 Saint Petersburg, Russia}
    \affiliation{Rzhanov Institute of Semiconductor Physics, Siberian Branch, Russian Academy of Sciences, 630090 Novosibirsk, Russia}
\author{T.~Iwata}
    \affiliation{Graduate School of Advanced Science and Engineering, Hiroshima University, 739-8526 Higashi-Hiroshima, Japan}
    \affiliation{International Institute for Sustainability with Knotted Chiral Meta Matter $(\text{WPI-SKCM}^2)$, Hiroshima University, 739-8526 Higashi-Hiroshima, Japan}
\author{K.~Kuroda}
    \affiliation{Graduate School of Advanced Science and Engineering, Hiroshima University, 739-8526 Higashi-Hiroshima, Japan}
    \affiliation{International Institute for Sustainability with Knotted Chiral Meta Matter $(\text{WPI-SKCM}^2)$, Hiroshima University, 739-8526 Higashi-Hiroshima, Japan}
    \affiliation{Research Institute for Semiconductor Engineering, Hiroshima University, 739-8527 Higashi-Hiroshima, Japan}
\author{K.~Miyamoto}
    \affiliation{Research Institute for Synchrotron Radiation Science, Hiroshima University, 739-0046 Higashi-Hiroshima, Japan}
\author{T.~Okuda}
    \affiliation{International Institute for Sustainability with Knotted Chiral Meta Matter $(\text{WPI-SKCM}^2)$, Hiroshima University, 739-8526 Higashi-Hiroshima, Japan}
    \affiliation{Research Institute for Semiconductor Engineering, Hiroshima University, 739-8527 Higashi-Hiroshima, Japan}
    \affiliation{Research Institute for Synchrotron Radiation Science, Hiroshima University, 739-0046 Higashi-Hiroshima, Japan}
\author{K.~Shimada}
    \affiliation{Graduate School of Advanced Science and Engineering, Hiroshima University, 739-8526 Higashi-Hiroshima, Japan}
    \affiliation{International Institute for Sustainability with Knotted Chiral Meta Matter $(\text{WPI-SKCM}^2)$, Hiroshima University, 739-8526 Higashi-Hiroshima, Japan}
    \affiliation{Research Institute for Semiconductor Engineering, Hiroshima University, 739-8527 Higashi-Hiroshima, Japan}
    \affiliation{Research Institute for Synchrotron Radiation Science, Hiroshima University, 739-0046 Higashi-Hiroshima, Japan}
\author{A.V.~Tarasov}
    \affiliation{Saint Petersburg State University, 198504 Saint Petersburg, Russia}

\begin{abstract}

    The spin-resolved dispersion dependencies for the topological insulator \mgbt{}
    in the {\kgk} path of the Brillouin zone (BZ) were studied by spin- and
    angle-resolved photoemission spectroscopy using laser radiation (Laser
    Spin-ARPES) with variation of the concentration of substitutional Ge atoms ($x$
    from 0.1 to 0.8) for in-plane ($s_x$) and out-of-plane ($s_z$) spin orientation.
    The formation of Rashba-like states is shown, which shift to lower energies with
    increasing Ge concentration. In the region of Ge concentrations from 50\% to
    75\%, the contribution of these states to the formed spin-dependent dispersions
    becomes predominant. A pronounced in-plane ($s_y$) spin polarization, asymmetric
    for opposite $\pm k_\parallel$ directions, is revealed for the Dirac cone
    states, while the Rashba-like states exhibit a pronounced asymmetry in both
    in-plane ($s_y$) and out-of-plane ($s_z$) spin polarizations. Theoretical
    calculations confirmed the asymmetric polarization of the Rashba-like states
    formed in the {\kgk} path of the BZ, simultaneously for in-plane and
    out-of-plane spin orientation. Constant energy maps for Rashba-like states show
    a pronounced $s_z$ spin component along the {\gk} direction, with a sign change
    as the contour crosses the {\gm} direction. The observed spin polarization can
    influence the development of spin devices based on magnetic topological
    insulators.

\end{abstract}

\maketitle

\label{sec:level1}
\section{Introduction}

The interplay of magnetism and band topology in magnetic topological insulators
(TIs) leads to many exotic quantum phenomena, for example, quantized anomalous
Hall effect (QAHE), topological magnetoelectric effect (TME) and other unique
topological properties \cite{qi2008topological, qi2011topological,
tokura2019magnetic, chang2023colloquium, chang2013experimental,
wang2021intrinsic}. Intrinsic magnetic topological insulators are a class of
materials where magnetic atoms have definite crystallographic positions, the
most notable representative being \mbt{} which was synthesized and
experimentally studied several years ago \cite{otrokov2019prediction,
li2019intrinsic, zhang2019topological, gong2019experimental}. Since magnetic
atoms are arranged regularly, quantum topological effects such as QAHE can be
observed at significantly higher temperatures, for example, QAHE is reachable in
\mbt{} thin films at 1.4~K without any external magnetic field or at 6.5~K when
such field is applied \cite{deng2020quantum}. 

\mbt{} is an antiferromagnetic (AFM) TI with a N\'{e}el temperature of
approximately 24---25~K \cite{otrokov2019prediction, li2019intrinsic,
zhang2019topological, gong2019experimental, deng2020quantum}. Experimental
studies of its surface electronic structure revealed an anomalously large energy
gap up to 90~meV \cite{otrokov2019prediction, li2019intrinsic,
zhang2019topological, gong2019experimental, deng2020quantum,
estyunin2020signatures, shikin2021sample, shikin2020nature, shikin2023routes,
garnica2022native, shikin2023topological} in its topological surface states
(TSS) which, however, exhibits such significant variations that even a gapless
surface spectrum may be observed \cite{hao2019gapless, chen2019topological,
swatek2020gapless}.  Numerous theoretical studies of \mbt{}-family materials
reveal their band structures to be particularly sensitive to slight
modifications of their geometry (e.g. first van der Waals gap magnitude), local
atomic composition (lower-$Z$ dopants of magnetic moments or nonmagnetic doping
of the magnetic layers) or other factors which influence TSS localization depth
relative to the crystal surface \cite{shikin2021sample, shikin2020nature,
shikin2023routes, garnica2022native, shikin2023topological,
chowdhury2019prediction, zhang2021tunable, zhou2020topological,
shikin2022modulation, shikin2024study}.

Such sensitive TSS behavior suggests that topological phase transitions (TPTs)
may be driven externally in these materials \cite{shikin2021sample,
shikin2020nature, shikin2023routes, garnica2022native, shikin2023topological}.
It is hypothesized that even an axion-like state with its corresponding TME
effect may be observed near the TPT point \cite{shikin2023topological}, which
can be driven by application of an external electric field
\cite{shikin2022modulation, shikin2024study}. Another possibility to influence
the TPT can be a doping the Mn layer with atoms of IV group such as Ge, Sn or Pb
\cite{tarasov2023topological, estyunina2023evolution, frolov2024magnetic,
qian2022magnetic, changdar2023nonmagnetic, zhu2021magnetic, yan2021elusive}. The
resulting materials have the \mabt{} stoichiometry ($A$ = Ge, Sn, Pb) where TPTs
are likely to occur for $x \approx 40\%\ldots50\%$ \cite{tarasov2023topological,
estyunina2023evolution, frolov2024magnetic}, even a Weyl semimetal state is
suggested for certain combinations of magnetic and atomic structures
\cite{shikin2024phase}. Characteristic N\'{e}el temperatures and spin-flop
transition field threshold for these compounds decrease from 24.5~K and 7~T in
pristine \mbt{} to $10\ldots15$~K and $1\ldots1.5$~T, respectively, when $x$
reaches $\approx 50\%\ldots80\%$; further doping leads to values of 5~K and
0.5~T, respectively \cite{estyunin2023comparative}.

Due to the unique spin-momentum locked spin structure and corresponding
charge-spin interconversion effects, magnetic TIs (including the \mbt{} family
\cite{shikin2020nature}) are now increasingly being used in spintronics
\cite{zhang2016conversion}. For instance, the injection of an electric current
into a TI is known to generate in-plane spin-transfer torques (STTs) orders of
magnitude larger than in other materials with high spin-orbit coupling (SOC)
strength \cite{mellnik2014spin}. These giant STTs are suitable for low-power
current-induced magnetization switching in heterostructures consisting of a TI
with a ferromagnetic (FM) layer.

Moreover, the design of topological insulators with out-of-plane surface spin
texture could further expand the range of spintronic effects and related
applications \cite{han2021topological, liu2020two}. For example, the
current-induced magnetization of layers with perpendicular magnetic anisotropy,
which are typically used in logic and memory devices, is switched more
efficiently using the out-of-plane STTs instead of in-plane STTs
\cite{ikeda2010perpendicular}.

In this study we explore Ge concentration-dependent in-plane and out-of-plane
spin structure of \mgbt{} experimentally (through Laser Spin ARPES) and
theoretically (using density functional theory (DFT) calculations), examining
the underlying mechanisms driving  the features of spin structure and their
evolution.

\section{Results of experimental measurements}

\begin{figure*}[ht!]
    \centering
    \includegraphics[width=14cm]{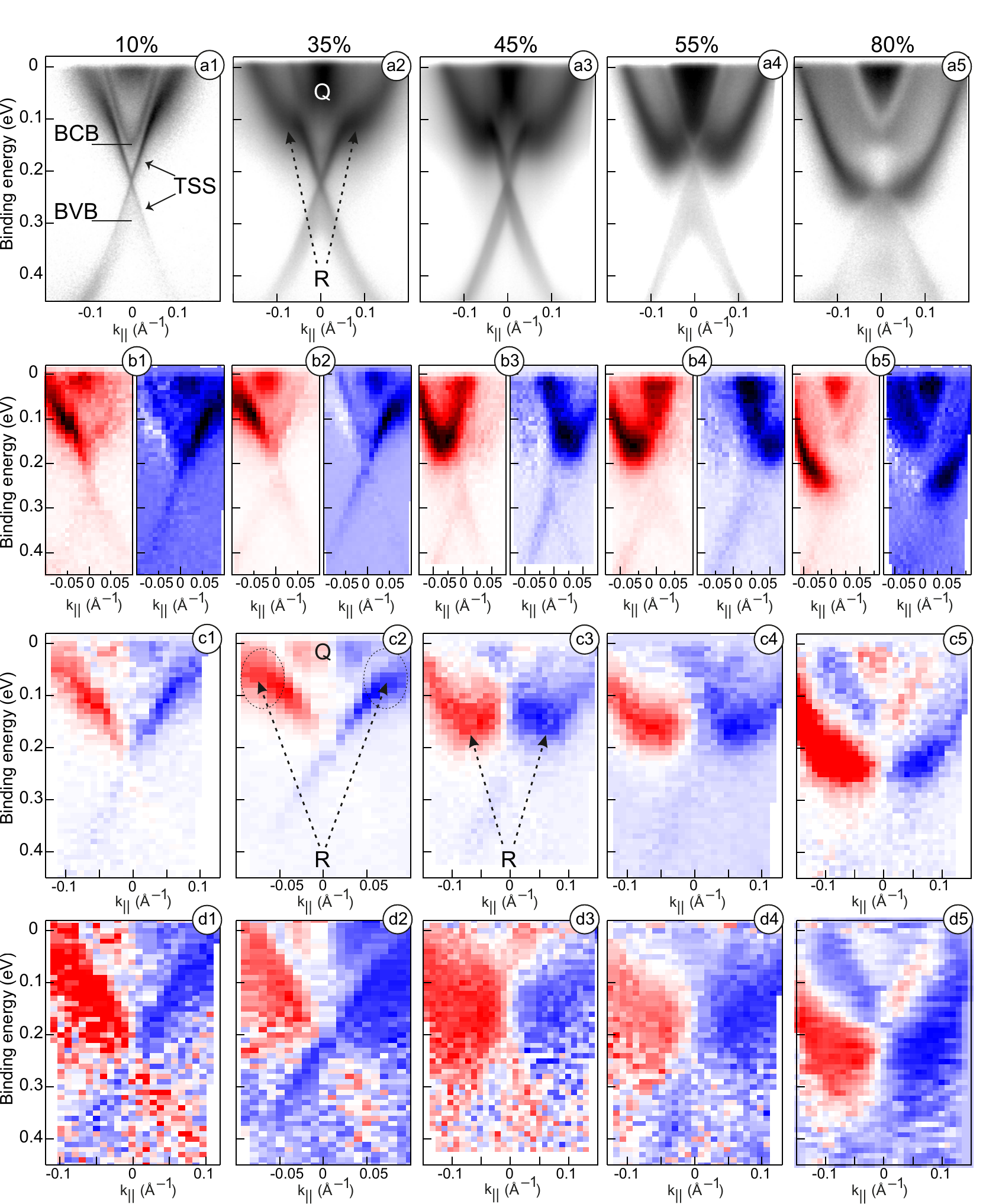}

    \caption{\textbf{(a1--a5)} Experimental spin-integrated ARPES data for
    \mgbt{} with different $x$ measured using $s$-polarized laser radiation with
    photon energy $h\nu = 6.3$~eV. \textbf{(b1--b5)} Experimental
    spin-resolved ARPES data for in-plane spin polarization with opposite spin
    orientations ($I_\uparrow$ in red, $I_\downarrow$ in blue) reconstructed
    using the Sherman function. \textbf{(c1--c5)} Intensity-normalized maps of
    spin polarization difference ($I_\uparrow - I_\downarrow$) where dominant
    polarizations are color-coded according to the panel (\textbf{b}).
    \textbf{(d1--d5)} Conventional spin polarization maps calculated from raw
    data by the formula $(I_\uparrow-I_\downarrow)/(I_\uparrow +
    I_\downarrow)$.}

    \label{fig1}
\end{figure*}

\begin{figure*}[ht!]
    \centering
    \includegraphics[width=16cm]{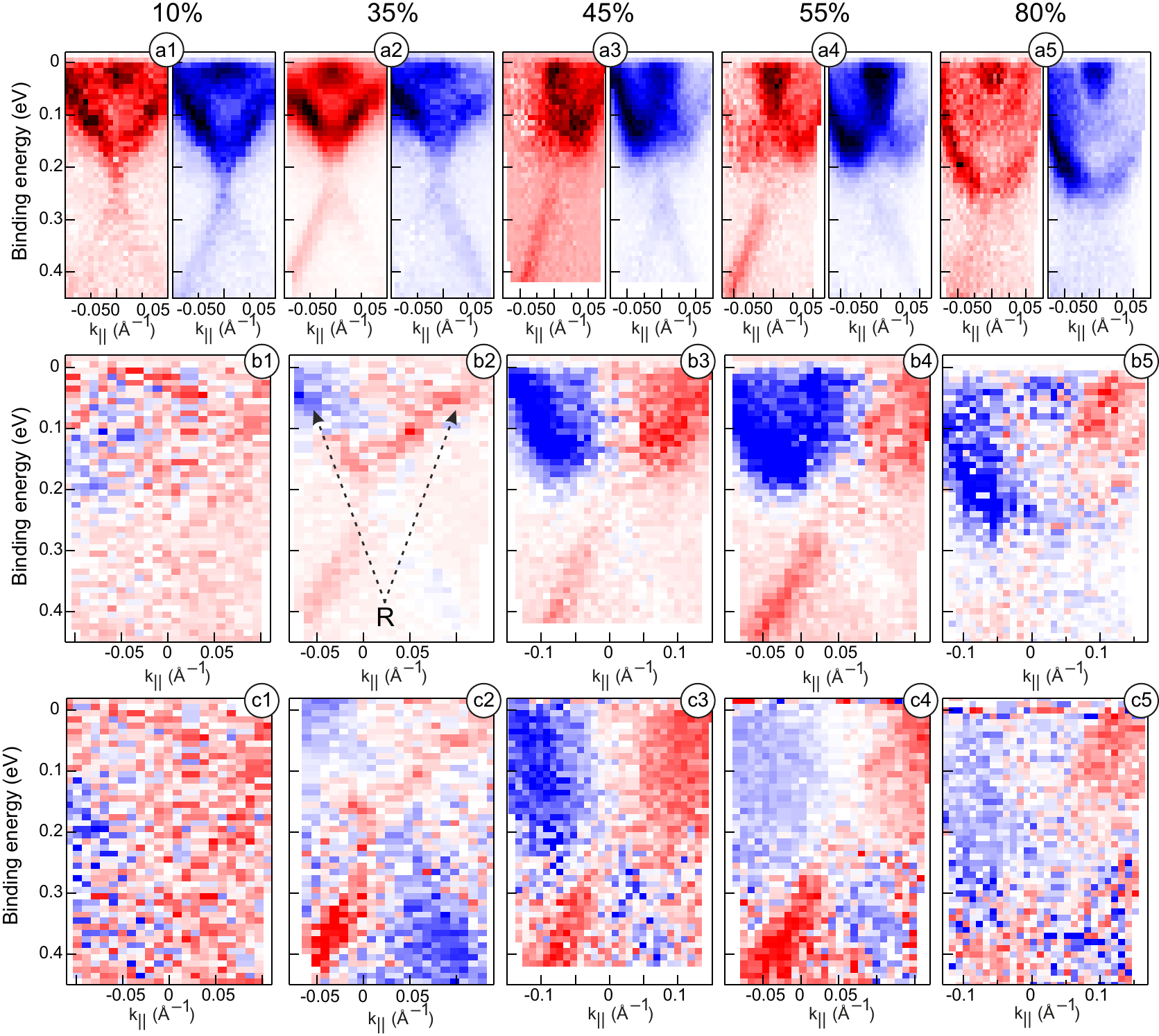}

    \caption{\textbf{(a1--a5)} Experimental spin-resolved ARPES data for
    out-of-plane spin polarization with opposite spin orientations ($I_\uparrow$
    in red, $I_\downarrow$ in blue) reconstructed using the Sherman function.
    \textbf{(b1--b5)} Intensity-normalized maps of spin polarization difference
    ($I_\uparrow - I_\downarrow$) where dominant polarizations are color-coded
    according to the panel (\textbf{a}). \textbf{(c1--c5)} Conventional spin
    polarization maps calculated from raw data by the formula
    $(I_\uparrow-I_\downarrow)/(I_\uparrow + I_\downarrow)$.}

    \label{fig2}
\end{figure*}

Spin-resolved ARPES measurements of \mgbt{} for Ge concentration values $x$ of
10\%, 35\%, 45\%, 55\% and 80\% were performed using $s$-polarized laser
radiation with photon energy $h\nu = 6.3$~eV for both in-plane (Fig.~\ref{fig1})
and out-of-plane (Fig.~\ref{fig2}) spin polarizations. Fig.~\ref{fig1}(b) and
Fig.~\ref{fig2}(a) show spin-resolved data with Sherman function corrections for
spin-up ($I_\uparrow$, red) and spin-down ($I_\downarrow$, blue) orientations.
Fig.~\ref{fig1}(c) and Fig.~\ref{fig2}(b) present the normalized maximal
intensity differences $I_\uparrow - I_\downarrow$, while Fig.~\ref{fig1}(d) and
Fig.~\ref{fig2}(c) show spin polarization diagrams calculated as
$\frac{I_\uparrow - I_\downarrow}{I_\uparrow + I_\downarrow}$. Experimental band
structures without spin resolution are shown in Fig.~\ref{fig1}(a) as a
reference for assessing band structure evolution with increasing Ge
concentration, similar to earlier works \cite{estyunina2023evolution,
shikin2024phase}.

\paragraph*{Spin-integrated band structure.} At 10\% doping, the band structure
in Fig.~\ref{fig1}(a1) shows typical TSS Dirac cone structure with Dirac point
(DP) position located at the energy of 0.22~eV. Valence and conduction band (VB
and CB) edges are observed at the energy of 0.30~eV and 0.15~eV, respectively. A
feature marked Q appears near the Fermi level (FL) with further development at
higher Ge-concentrations (it is more visible in spin-resolved dispersion).


With increasing Ge concentration, a new feature (marked R) appears in the CB
energy region, attributed to Rashba-like states with surface character
\cite{shikin2024phase}. These states evolve with Ge doping and are
Ge-contributed.  At 35\% Ge and then at 45\% Ge (Fig.~\ref{fig1}(a2, a3)),
Rashba-like states coexist with the Dirac cone, but further concentration
increase makes original TSS indistinguishable whereas these Rashba-like states
shift in energy and become a prevalent band structure feature along with gapless
bulk states. At 55\% Ge (Fig.~\ref{fig1}(a4)), Rashba-like states form two
distinct parabolic branches split by CB states at the $\Gamma$~point. The outer
branches of these states (feature R) shift to lower energies as $x$ increases,
reaching the energy of 0.25~eV at 80\% Ge (Fig.~\ref{fig1}(a5)), surpassing the
original DP position. At such high concentrations, the bulk band gap reopens,
and the band structure resembles Mn-doped \gbt{}.  Similar Rashba-like states
have been previously observed in \mgbt{} and \msbt{} systems
\cite{tarasov2023topological, estyunina2023evolution, shikin2024phase},
suggesting that the energy shifts observed in our measurements may be a common
feature across these materials.

\paragraph*{Spin-resolved band structure.} Spin-resolved ARPES reveals detailed
band structure features. At 10\% doping, the in-plane spin structure of the
Dirac cone shows typical spin-momentum locking for helical TSS in topological
insulators, as seen in Fig.~\ref{fig1}(d1). The in-plane spin polarization
reverses ($I_\uparrow \longleftrightarrow I_\downarrow$) together with planar
electron quasimomentum ($-k_{\parallel} \longleftrightarrow +k_\parallel$). 

The studied system with Ge concentration of 35\% has a very similar in-plane
spin structure in Fig.~\ref{fig1}(b2---d2) which exhibits in-plane spin reversal
when the planar quasimomentum $k_\parallel$ is reversed. In particular, the left
branches of both  features R (Rashba-like states) and Q are positively polarized
($I_\uparrow$). The out-of-plane spin structure becomes more noticeable in
Fig.~\ref{fig2}(c2), where the lower Dirac cone part acquires detectable
out-of-plane polarization ($I_\uparrow$ for left branch, $I_\downarrow$ for
right branch). The above-mentioned Rashba-like states with opposite out-of-plane
polarizations begin to manifest themselves already at 35\% Ge near the Fermi
level (marked as R states with corresponding arrows), see Fig.~\ref{fig2}(b2).

At 45\% and higher Ge concentrations, the spin polarization of feature Q is
reversed at the same time when parabolic branches of Rashba-like states (R)
become pronounced. This supports the idea that feature Q is a continuation of
the branches of the Rashba-like states. Both in-plane and out-of-plane
polarizations of feature R intensify with increasing Ge content, while feature Q
shows also the out-of-plane polarization.  At higher Ge-concentrations ($x = 80\%$) 
the Rashba-like states still exhibit their pronounced character.

\section{Results of theoretical calculations}

\begin{figure*}
\centering
    \includegraphics[width=16cm]{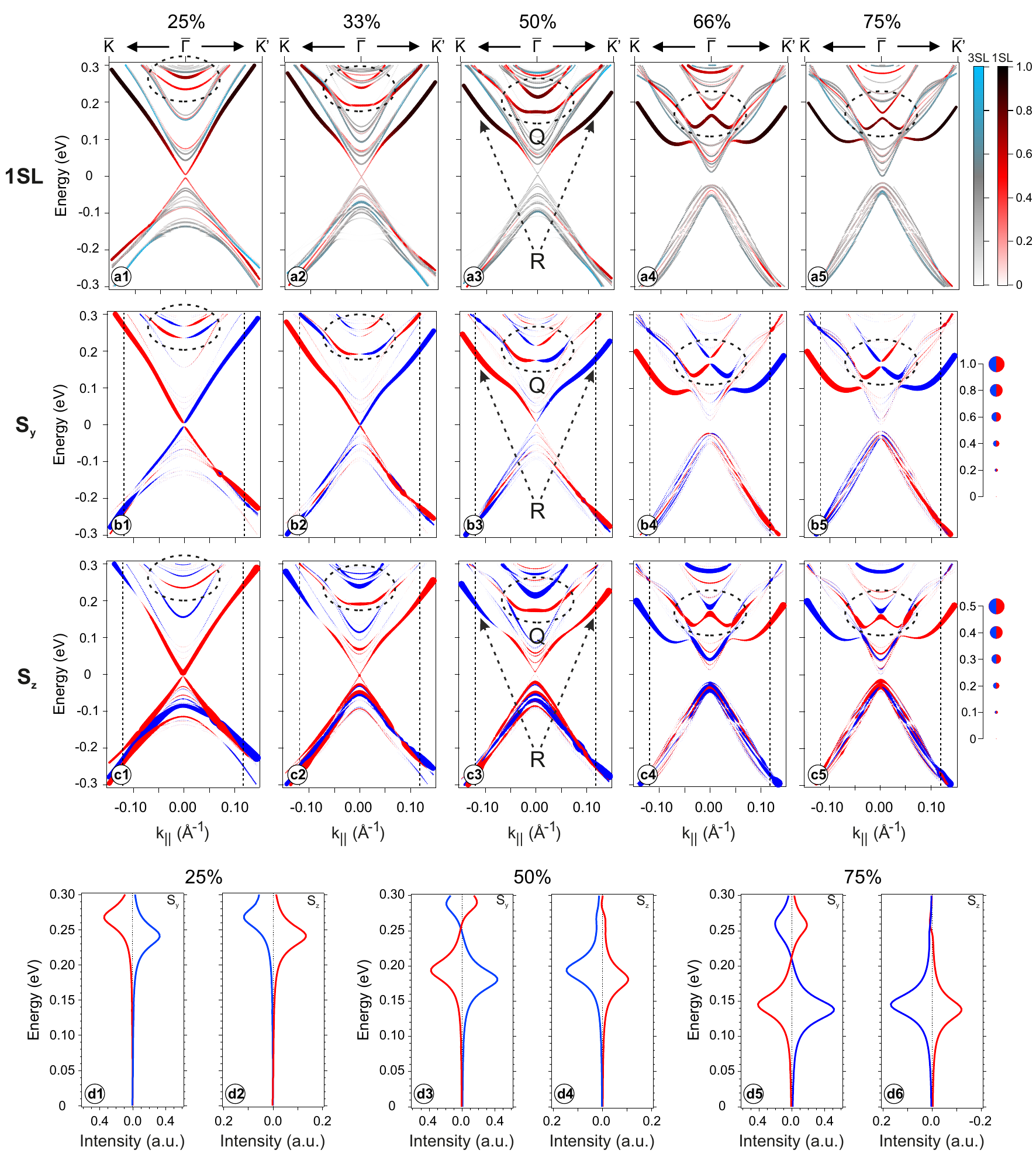}

    \caption{\textbf{(a1-a5)} Calculated band structure of 12 SL thick slab of
    \mgbt{} along the {\gk} direction ($x$) in the BZ for Ge concentrations from
    25\% to 75\% with distinguished the first (white-red-black) and third
    (gray-blue) SLs contributions, color intensity defines the localization
    degree. \textbf{(b1--b5)} and \textbf{(c1--c5)} depict in-plane $s_y$ and
    out-of-plane $s_z$ spin polarizations, respectively.  Positive and negative
    spin polarizations along the corresponding directions are shown by red and
    blue symbols, where symbol sizes determine the spin polarization magnitudes.
    \textbf{(d1--d6)} Spin distribution profiles for $x$ values of 25\%, 50\%
    and 75\% taken along vertical dashed lines at $k_\parallel = \pm
    0.12$~\AA$^{-1}$ for bands with $s_y$ and $s_z$ spin polarizations.}

    \label{fig3} 
\end{figure*}

In order to understand the band and spin structure of \mgbt{}, a DFT
computational study was performed on supercells to simulate Ge doping (see
Methods for details). Fig.~\ref{fig3} shows the calculated band structures for
Ge concentrations of 25\%, 33\%, 50\%, 66\%, and 75\% along the {\gk} direction
($x$ axis). Panel \ref{fig3}\textbf{(a)} illustrates spatial localizations of
the states, while panels \ref{fig3}\textbf{(b)} and \ref{fig3}\textbf{(c)} show
$s_y$ (in-plane, perpendicular to the Brillouin zone path) and $s_z$
(out-of-plane) spin polarizations, respectively.

According to the obtained results, the calculated bands display behavior very
similar to the experimental data. Increasing Ge concentration leads to the
formation of Rashba-like states (features R and Q) with significant in-plane (up
to 50\%, see panels \ref{fig3}(\textbf{b},\textbf{d})) and out-of-plane spin
polarizations (up to 15\%, see panels \ref{fig3}(\textbf{c},\textbf{d})). These
states (R) become more pronounced and shift to lower energies, consistent with
previous DFT studies of similar systems doped with Ge or Sn
\cite{tarasov2023topological, estyunina2023evolution, shikin2024phase}. 

Moreover, these Rashba-like states are almost completely localized at the
surface in the first SL (see panel \ref{fig3}(\textbf{a})) for all Ge
concentrations, while the TSS, which exist at low Ge concentrations, diffuse
deeper into the bulk. Notably, at $x = 25\%$, the upper Dirac cone band already
shows strong surface localization in the energy range $+0.2\ldots0.3$~eV (see
Fig.~\ref{fig3}(a1)), where the Rashba-like states begin to form as Ge doping
increases.

Fig.~\ref{fig3}(d1--d6) shows spin profiles for Ge concentrations of 25\%, 50\%,
and 75\% along the vertical dashed lines $k_\parallel = \pm 0.12$~\AA$^{-1}$,
illustrating the in-plane and out-of-plane spin polarizations of the intersected
states (mostly the outer branches of the Rashba-like states). It is clearly
observed that the out-of-plane polarization is actually much weaker than the
in-plane polarization.

\paragraph*{In-plane spin structure.} At 25\% Ge the band structure shows
spin-momentum locking typical for TSS (Fig.~\ref{fig3}(b1)) while another band
located at $+0.25$~eV displays similar behavior. This band likely represents
feature Q from the experimental spectra in Fig.~\ref{fig1}(a1--d1). As Ge
concentration increases, the TSS disappear ($x = 50\%$), with bands below the DP
losing in-plane spin polarization and the Rashba-like states above the DP
gaining it. By 66\% Ge, the parabolic Rashba-like bands become pronounced, and
features R and Q shift down in energy, consistent with experimental results
(Fig.~\ref{fig1}(c3--c5)).


\paragraph*{Out-of-plane spin structure.} Fig.~\ref{fig3}(\textbf{c}) shows three
notable observations. First, at low Ge concentrations, spin-polarized features
appear at $+0.25$~eV (which later form the Rashba-like states), as also seen in
the experimental data (Fig.~\ref{fig2}(b2)). At $x = 66\%$ and 75\%,
well-defined Rashba-like states are formed, with out-of-plane spin polarization
also inverted relative to the $\Gamma$ point. Only the outer branches of the
Rashba-like states (feature R) show out-of-plane spin inversion with
$k_\parallel$ reversal, clearly seen experimentally in Fig.~\ref{fig2}(b3, b4).
In these outer branches, both red and blue regions are present, indicating
opposite spin projections for different states as the wave vector $k_\parallel$
increases. The feature Q and other states show consistent polarization
regardless of quasimomentum direction, as confirmed by experimental features
near the FL (Fig.~\ref{fig2}(a4--a5)).

\begin{figure*}
\centering
    \includegraphics[width=16cm]{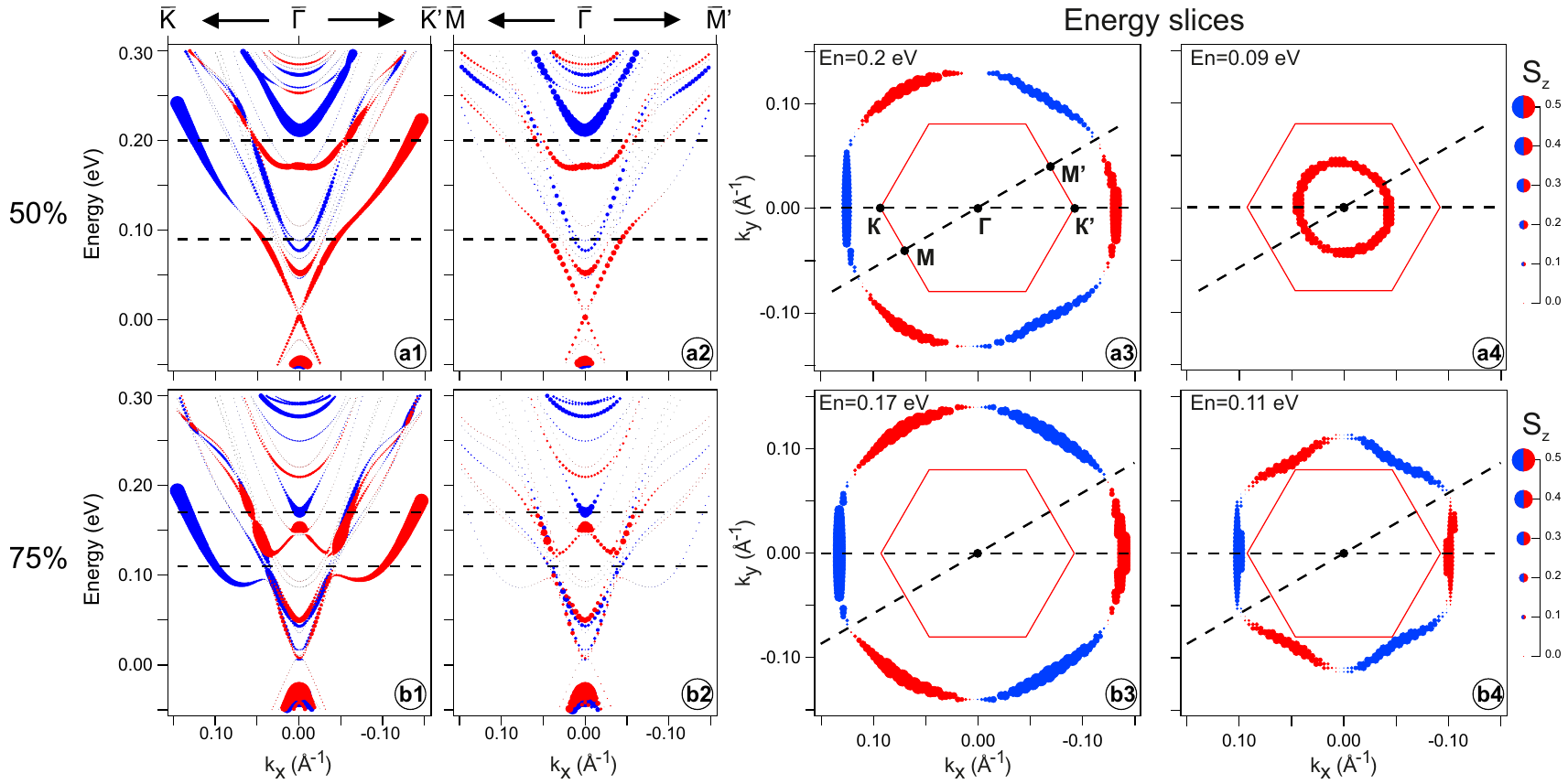} 

    \caption{Constant energy contours for outer branches of the Rashba-like
    states with out-of-plane spin polarization for \mgbt{} with 50\% and 75\% Ge
    concentrations. Spin-resolved dispersions along the {\gk} and {\gm}
    directions together with out-of-plane spin polarization are provided for
    reference; the corresponding energy levels are shown with dashed lines. Each
    energy contour is provided together with a hexagonal image of the Brillouin
    zone (not to scale) which displays alignment of principal symmetry
    directions.}

    \label{fig5}
\end{figure*}

Further exploration of the non-trivial out-of-plane spin texture is possible
through out-of-plane spin-polarized constant energy contours in the $k_x$--$k_y$
plane. These contours, shown in Fig.~\ref{fig5}, depict systems with $x = 50\%$
and 75\%. The constant energy contours taken far from the $\Gamma$ point exhibit
a trigonal spin texture, with the out-of-plane spin polarization alternating
every $60^\circ$ with maxima in the {\gk} direction and reaching zero at the
{\gm} direction. However, the images lack threefold rotation symmetry due to
broken inversion symmetry (Fig.~\ref{fig5}\textbf{(a4)}), where the contour
closer to $\Gamma$ shows positive out-of-plane spin polarization but is not
perfectly circular.

As a result, it can be stated that both experimental data and DFT calculations
for \mgbt{} confirm existence of surface states with parabolic-like outer
branches with helical in-plane spin texture as well as non-trivial out-of-plane
spin texture. In-plane helicity together with predominant surface localization
suggest strong resemblance of these states to those which occur in the canonical
Rashba effect with two-dimensional electron gas \cite{bychkov1984properties,
bihlmayer2022rashba, krasovskii2015spin}, hence the term \enquote{Rashba-like
states} used throughout this paper.

\section{Discussion}

Contrary to the simplest interpretation of the Rashba model
\cite{bychkov1984properties,krasovskii2015spin,ast2007giant,premper2007spin}, in
our case, the formation of Rashba-like spin-momentum locking states can be
associated with the asymmetric hybridization of $p_{x(y)}$ and $p_z$ orbitals,
mediated by $s$ orbitals, under interaction of Ge atoms (replacing Mn atoms)
with Te and Bi atoms in the nearest layers. This leads to the formation of a
$\mathbf k$-antisymmetric momentum-dependent orbital momentum $\mathbf L_n
(\mathbf k) = - \mathbf L_n (-\mathbf k)$, resulting from inversion symmetry
breaking \cite{bihlmayer2022rashba}. This orbital momentum couples to the spin
in the presence of spin–orbit coupling, forming Rashba-like spin-momentum
locking \cite{bihlmayer2022rashba,park2011orbital,go2017toward}. The formation
of such states depends on chemical bonds, making them Ge-contributed and
shifting in energy with changing Ge concentration. The influence of this orbital
momentum is strongest near the nucleus
\cite{bihlmayer2022rashba,krasovskii2015spin,park2011orbital}, where spin-orbit
coupling is most effective. The inclusion of Ge $p$ states (instead of Mn $sd$
states) enhances Rashba-like effects, since $p$-electrons approach the nucleus
more closely than $d$-electrons \cite{bihlmayer2022rashba,krasovskii2015spin}.
As a result, Rashba-like effects in systems based on $p$-electrons, such as
Ge-doped materials, are more pronounced, with the inversion symmetry breaking at
the surface leading to surface Rashba states, similar to classical Rashba
effects.

The formation of Rashba-like states with out-of-plane spin polarization ($s_z$)
was also observed in Bi/Ag systems \cite{ast2007giant, premper2007spin,
bihlmayer2007enhanced}. In \cite{meier2008quantitative}, it was shown that for
Bi/Ag(111) and Pb/Ag(111), an out-of-plane spin component ($s_z$) accompanies
in-plane polarization ($s_{x,y}$) due to partial spin rotation between the K and
M points of the surface BZ. This component is maximal along {\gk} and minimal
along {\gm}.

Similar effects of spin polarization transformation between $s_{x,y}$ and $s_z$
are seen in topological surface states (TSS) of topological insulators near the
Fermi level due to warping effects \cite{wang2011observation,
bahramy2012emergent, fu2009hexagonal, basak2011spin}.  These effects, which
create an out-of-plane spin component along {\gk} while preserving in-plane
polarization along {\gm}, are strongest near the Fermi level and minimal near
the Dirac point.

For the systems in this work, inversion symmetry breaking in the {\gk} direction
plays a significant role. Similar effects were observed in Refs
\cite{otrokov2019prediction, otrokov2017magnetic, tan2022momentum} for systems
based on \mbt{}, \gbt{}, \mbt/$\mathrm{Bi}_2\mathrm{Te}_3$ as well as Bi/\mbt{}
and Bi/$\mathrm{Bi}_2\mathrm{Se}_3$ bilayers \cite{sheverdyaeva2023giant,
klimovskikh2024interfacing}, where asymmetry in the {\gk} direction was linked
to crystal symmetry and the interplay of spin-orbit coupling and exchange
fields. As a result, the right and left branches of states interact with the
Zeeman field in different ways, leading to a difference in the left and right
branches of states ($E(+k)$ is not equal to $E(-k)$) and the formation of the
$s_z$ spin component in the {\kgk} path, which coexists with the in-plane
$s_{x,y}$ polarizations. Time-reversal and inversion symmetry breaking cause
$E(+k) \neq E(-k)$, leading to an $s_z$ spin component in the {\gk} direction
along with in-plane polarization \cite{otrokov2019prediction,
otrokov2017magnetic, tan2022momentum}. This results in a threefold, rather than
sixfold, momentum inversion asymmetry.

In contrast, along the {\gm} direction, the (0001) surface of \mbt{} has three
mirror planes, maintaining symmetry along {\gm}, which suppresses inversion
asymmetry and causes the $s_z$ component to approach zero in this direction
\cite{otrokov2019prediction, otrokov2017magnetic, tan2022momentum}.

To confirm this, Fig.~\ref{fig5} presents constant energy contours including
out-of-plane spin component for systems with 50\% and 75\% Ge concentrations.
Spin-resolved dispersion dependencies in the {\gk} and {\gm} directions for
energies above the DP are also shown which indicate the contour energy levels by
horizontal dotted lines. The contours taken at energies between 0.1 and 0.2~eV
show threefold symmetry with spin inversion in the $\pm k_\parallel$ Rashba-like
branches along the {\kgk} direction. This correlates with the dispersion
calculations (Fig.~\ref{fig3}) and experimentally measured dispersions
(Figs~\ref{fig1},~\ref{fig2}). Contours taken near 0.08--0.09~eV correspond to
the Dirac cone states and represent distorted circles without apparent spin
orientation changes in agreement with Fig.~\ref{fig3}. Along the {\gm}
direction, the spin inversion changes sign and $s_z$ component vanishes.

Thus, both experimental and theoretical results show good agreement for spin
orientations in the {\gk} direction, confirming the analysis of the observed
phenomena.

\section{Methods}

First-principles calculations in the framework of the density functional theory
(DFT) were performed at the Computing Center of SPbU Research park. The
electronic structure supercell calculations with impurities were performed using
the OpenMX DFT code which implements a linear combination of pseudo-atomic
orbitals (LCPAO) approach \cite{ozaki2003variationally, ozaki2004numerical,
ozaki2005efficient} with full-relativistic norm-conserving pseudopotentials
\cite{troullier1991efficient}. The GGA-PBE exchange-correlation functional
\cite{perdew1996generalized} was utilized and basis sets were specified as
Ge7.0-s3p2d2, Mn6.0-s3p2d1, Te7.0-s3p2d2f1 and Bi8.0-s3p2d2f1 where numbers mean
interaction ranges in \AA. Real-space numerical integration accuracy was
specified by cutoff energy of 200~Ry, total energy convergence criterion was
$2\times 10^{-5}$ Hartree. The Mn~$3d$ states were considered within the
$\text{DFT}+U$ approach \cite{dudarev1998cj} with $U = 5.4$~eV.

Simulation of \mgbt{} compounds for $x$ values of 25\%, 50\% and 75\% was
performed using $2 \times 2$ supercells whereas $3 \times 1$ supercells were
used for Ge concentrations of 33\% and 66\%. Surface band structures were
calculated for 12~SL slabs separated by 20~\AA{} vacuum gaps in the $z$
direction (the surface normal) while constant energy contours were calculated
using 6~SL slab thickness. Surface Brillouin zones of $2 \times 2$ ($3 \times
1$) supercells were sampled with regular $3 \times 3$ ($3 \times 5$) meshes of
$\mathbf k$-points, respectively. 

Spin-resolved APRES experiments were performed with LaserSpin ARPES station with
Scienta DA30 analyzer stationed at HiSOR \cite{iwata2024hisor} (Hiroshima,
Japan) with photon energy of $h\nu=6.3$~eV was used. All measurements were
conducted under ultra-high vacuum conditions (base pressure was less than $10
\times 10^{-10}$ Torr) and sample temperatures below 30~K. Stoichiometric
compositions of analyzed samples were estimated using X-ray photoelectron
spectroscopy (XPS) with Al~$K_{\alpha}$ ($h\nu=1486.7$~eV) radiation.

\section{Conclusion}

Experimental and theoretical studies of the spin electron structure with
in-plane and out-of-plane spin orientation for TI \mgbt{}~with a change in the
concentration of substituting Ge atoms are carried out. The formation of
Ge-contributed Rashba-like states is shown, characterized in addition to the
bulk-derived origin also by a significant surface contribution, which shift to
lower energies as the Ge concentration increases. These states determine the
main contribution to the formed structure of \mgbt.

Experimental studies and theoretical calculations have shown that these
Rashba-like states are characterized by pronounced in-plane and out-of-plane
spin polarizations along the {\gk} direction of the BZ, with inversion in the
$\pm k_\parallel$ directions relative to the $\Gamma$ point. Overall, a strong
correlation is observed between the experimentally measured spin-resolved
dispersion relations and the corresponding theoretical calculations. Theoretical
studies of constant energy maps for Rashba-like states also show that the sign
of the $s_z$ spin component changes as the contour crosses the {\gm} direction.

The observed spin polarization, especially the distinct in-plane and
out-of-plane spin textures of the Rashba-like states, opens up new opportunities
for the development of advanced spintronic devices. The strong spin-momentum
locking and the ability to control spin orientation via Ge doping make \mgbt{} a
promising candidate for applications in low-power spintronic technologies, such
as spin transistors and memory devices. Additionally, the tunable nature of the
Rashba-like states and their surface localization could be exploited in the
design of heterostructures, enabling efficient spin-charge interconversion and
current-induced magnetization switching. These properties position \mgbt{} as a
versatile material platform for future magnetic topological insulator-based
spintronic applications.

\begin{acknowledgments}
This work was supported by the Russian Science Foundation grant No. 23-12-00016
and the St. Petersburg State University grant No. 95442847. HiSOR ARPES
measurements were performed under Proposals No. 23AU012 and 23BU003. We are
grateful to the N-BARD, Hiroshima University for liquid He supplies.  The
calculations were partially performed using the equipment of the Joint
Supercomputer Center of the Russian Academy of Sciences
(https://rscgroup.ru/en/project/jscc).  Monocrystalline \mgbt{} samples were
grown using the Bridgeman method under the state assignment of Sobolev Institute
of Geology and Mineralogy SB RAS No. 122041400031-2.
\end{acknowledgments}

\end{document}